\documentclass[twocolumn,trackchanges]{aastex701}

\usepackage{makecell}
\usepackage{gensymb}
\usepackage{amsmath}
\usepackage{mathtools}
\usepackage{booktabs}

\newcommand{\vsini}{$v$\,sin\,$i_\star$}
\newcommand{\vsinimath}{v\,\text{sin}\,i_\star}
\newcommand{\vsinsin}{$\sqrt{v\,\text{sin}\,i_\star}$\,sin\,$\lambda$}
\newcommand{\vsincos}{$\sqrt{v\,\text{sin}\,i_\star}$\,cos\,$\lambda$}

\begin{document}

\title{The KPF-SLOPE Survey - Small, Compact Multi-Planet Systems Appear Spin-Orbit Aligned}

\author[0000-0002-9305-5101]{Luke B. Handley}
\altaffiliation{NSF Graduate Research Fellow}
\affiliation{Department of Astronomy, California Institute of Technology, Pasadena, CA 91125, USA}
\email[show]{lhandley@caltech.edu} 

\author[0000-0001-8638-0320]{Andrew W. Howard} 
\affiliation{Department of Astronomy, California Institute of Technology, Pasadena, CA 91125, USA}
\email{ahoward@caltech.edu}

\author[0000-0002-8958-0683]{Fei Dai} 
\affiliation{Institute for Astronomy, University of Hawai`i, 2680 Woodlawn Drive, Honolulu, HI 96822, USA}
\email{fdai@hawaii.edu}

\author[0000-0003-3856-3143]{Ryan A. Rubenzahl} 
\affiliation{Center for Computational Astrophysics, Flatiron Institute, 162 Fifth Avenue, New York, NY 10010, USA}
\email{rrubenzahl@flatironinstitute.org}

\author[0000-0002-8965-3969]{Steven Giacalone} 
\altaffiliation{NSF Astronomy and Astrophysics Postdoctoral Fellow}
\affiliation{Department of Astronomy, California Institute of Technology, Pasadena, CA 91125, USA}
\email{steveng@caltech.edu}

\author[0000-0002-0531-1073]{Howard Isaacson} 
\affiliation{{Department of Astronomy,  University of California Berkeley, Berkeley CA 94720, USA}}
\email{hisaacson@berkeley.edu}

\author[0000-0001-7664-648X]{J. M. Joel Ong} 
\affiliation{Sydney Institute for Astronomy (SIfA), School of Physics, University of Sydney, NSW 2006, Australia}
\email{joel.ong@sydney.edu.au}

\author[0000-0001-6416-1274, sname=Carmichael, gname=Theron]{Theron W. Carmichael} 
\altaffiliation{NSF Ascend Postdoctoral Fellow}
\affiliation{Institute for Astronomy, University of Hawai‘i, 2680 Woodlawn Drive, Honolulu, HI 96822, USA}
\email{tcarmich@hawaii.edu}


\author[0000-0003-3020-4437]{Yaguang Li} 
\affiliation{Institute for Astronomy, University of Hawai`i, 2680 Woodlawn Drive, Honolulu, HI 96822, USA}
\email{yaguangl@hawaii.edu}

\author[0000-0001-8342-7736]{Jack Lubin}
\affiliation{Department of Physics \& Astronomy, University of California Los Angeles, Los Angeles, CA 90095, USA}
\email{jblubin@ucla.edu}

\author[0000-0001-5728-4735]{Pranav H. Premnath} 
\affiliation{Department of Physics \& Astronomy, University of California Irvine, Irvine, CA 92697, USA}
\email{premnatp@uci.edu}

\author[0009-0005-5520-1648]{Claire J. Rogers} 
\affiliation{Department of Physics \& Astronomy, University of California Irvine, Irvine, CA 92697, USA}
\email{c.j.rogers@uci.edu}

\author[0000-0002-1386-0603]{Pranav Nagarajan} 
\affiliation{Department of Astronomy, California Institute of Technology, Pasadena, CA 91125, USA}
\email{pnagaraj@caltech.edu}

\author[0000-0003-0742-1660]{Gregory J. Gilbert} 
\affiliation{Department of Astronomy, California Institute of Technology, Pasadena, CA 91125, USA}
\email{ggilbert@caltech.edu}

\author[0000-0003-3504-5316]{Benjamin Fulton} 
\affiliation{NASA Exoplanet Science Institute / Caltech-IPAC,
Pasadena, CA 91125, USA}
\email{bjfulton@ipac.caltech.edu}



\author[0009-0004-4454-6053]{Steven R. Gibson} 
\affiliation{Caltech Optical Observatories, California Institute of Technology, Pasadena, CA 91125, USA}
\email{sgibson@caltech.edu}

\author[0000-0001-8127-5775]{Arpita Roy} 
\affiliation{Astrophysics \& Space Institute, Schmidt Sciences, New York, NY 10011, USA}
\email{arpita308@gmail.com}



\author[0009-0002-2419-8819]{Jerry Edelstein} 
\affiliation{Space Sciences Laboratory, University of California, Berkeley, CA 94720, USA}
\email{jerrye@ssl.berkeley.edu}

\author[0009-0004-7325-3591]{Christopher Smith} 
\affiliation{Space Sciences Laboratory, University of California, Berkeley, CA 94720, USA}
\email{christopher.smith@berkeley.edu}





\begin{abstract}

The angle between stellar spin axes and planetary orbits---stellar obliquity---probes the dynamics of planetary migration and evolution. The obliquities of giant planets have been extensively studied because they are the most easily measured. Smaller planets, while more difficult to measure, have the advantage of better reflecting the dynamics of planetary systems because they trigger negligible back-reactions onto the host star. This paper introduces a new observational campaign called the Small, Low-mass Oblique Planets Experiment (SLOPE) survey with the Keck Planet Finder (KPF) spectrograph, and presents four new obliquity measurements. The SLOPE survey focuses on planets smaller than Saturn across a variety of system architectures. The sky-projected obliquities of the four planets measured---TOI-1386~b, TOI-480~b, TOI-4596~b, and TOI-1823~b---are all consistent with spin-orbit alignment. We validate the planetary nature of TOI-4596~b with a significant obliquity detection. Including these measurements, we conducted a statistical analysis of the obliquities of sub-Saturn size planets in different planetary system architectures. Compared to other architectures, those in compact multi-planet systems reside in orbits that appear preferentially aligned with the stellar equator with 6$\,\sigma$ confidence.

\end{abstract}

\section{Introduction} \label{sec:intro}

The distribution of angular momenta in an exoplanetary system, i.e., the various spin and orbital axes of a star and its planets, carry the imprint of the system's history. The angular momenta are shaped by formational processes that set the initial conditions and longer timescale dynamics that evolve the system. The well-established physics of molecular cloud collapse \citep{Jeans1902,Larson1969,Hennebelle2012}, protoplanetary disk formation \citep{Mestel1965,Spitzer1978,Bodenheimer1995}, and planetary conglomeration \citep{Safronov1969,Goldreich1973,Chiang2010} suggest that closely aligned angular momenta are the expected (although perhaps not universal) starting point. Under this assumption, misalignments may be interpreted as signatures of past dynamical activity, and the details of misalignments observed in a large sample of systems may constrain specific physical mechanisms.

Although giant planets are the most accessible targets for angular-momentum measurements, their large masses can strongly couple to and reshape their environments through processes such as tidal realignment \citep{Winn2010b,WangX2026} and spin–orbit coupling \citep{Zanazzi2018}, leading to complex N-body interactions and interpretive degeneracies. Small planets---here defined as ranging from Earth-sized to sub-Saturns---more closely approximate idealised test particles that induce no back-reaction and trace only the local gravitational potential (e.g., the tidal damping time scales inversely proportional with planetary mass; \citealt{Albrecht2022}). The spin-orbit angles of these planets therefore serve as cleaner probes of the underlying dynamical environment.

Understanding those environments is critical, as they may substantially alter inherent planetary occurrence and composition. Notably, \cite{CastroGonzalez2024} recently discovered a relative overabundance of Neptune sized planets at orbital periods of $\sim$3-6 days. It is unclear whether that pileup is a common end-state for planets undergoing exotic, high eccentricity migration mechanisms \citep{Wu2003,Fabrycky2007,Wu2011,Naoz2011,Naoz2012}, a halting point in the quiescent migration of planets via their natal disks \citep{Goldreich1979,Goldreich1996,Nelsion2000,Tanaka2002,Paardekooper2010}, the location of a dramatic phase change in atmospheric structure triggered by proximity to the photoionizing host \citep{Owen2012, Owen2019}, or a combination of mechanisms \citep{Owen2018, Bourrier2018b, Attia2021, Bourrier2025}. The three-dimensional architectures of planetary systems measured through, e.g., the Rossiter McLaughlin effect (RM effect---which specifically measures the difference between stellar spin and planetary orbital angular momentum, hereafter the stellar obliquity angle; \citealt{Rossiter1924,McLaughlin1924}) may be the key to interpreting both the kinds of planets we observe and what orbits they occupy (\citealt{Fabrycky2009}, \citealt{Triaud2010}, \citealt{Winn2010b}, and \citealt{Albrecht2012} are important examples for hot Jupiters).

This paper is organized as follows. In Section \ref{sec:survey}, we introduce our survey design, discuss the systems presented, and recapitulate/refine their stellar properties. In Section \ref{sec:analysis}, we introduce the analysis pipeline and discuss our modeling strategy. In Section \ref{sec:results}, we present our constraints on the stellar obliquities and discuss the validation of the planet TOI-4596~b. In Section \ref{sec:ensemble}, we analyze the ensemble of small planets, suggesting the alignment of compact multi-planet systems.

\section{The SLOPE Survey} \label{sec:survey}

Constraining the spin-orbit angles of small planets is difficult given that the RM amplitude scales as the square of the planetary radius. A successful stellar obliquity measurement requires multiple exposures within the transit window (tens of exposures in a few hours). Increasing the integration time to improve radial velocity (RV) precision would incur a tradeoff in temporal resolution, and thus precision in the obliquity measurement. The Keck Planet Finder (KPF) on Keck I is one of the few facilities with the aperture, Doppler precision, and line profile stability required for reliable obliquity measurements of targets with small RM signatures.  KPF has demonstrated $<$ 30 cm s$^{-1}$\ precision on timescales of hours (during an exoplanet transit; e.g., \citealt{Li2025}), making it a premier instrument for these observations.

The SLOPE survey will leverage these attributes to measure the obliquities of small planets among a variety of system architectures (e.g., the number and spacing of the planets) and assess how past dynamical interactions could manifest in the distribution of orbital inclinations (i.e., tilted orbits).  A second focus of the survey is on young planets, which will be compared with mature systems to understand the timescales of the dynamical interactions, constraining the relevant theoretical models. Unraveling those dynamics will paint a clearer picture of the formational environment, enriching our understanding of planetary architectures.

This population of planets has relatively few obliquity measurements. To date, obliquity constraints have been reported for approximately 50 sub-Saturn–mass planets. A subset of these were obtained with KPF, including HD~191939~b, Kepler~1656~b, TOI-1694~b, TOI-2374~b, TOI-2364~b, TOI-880~c and TOI-4495~c \citep{Lubin2024,Rubenzahl2024a,Handley2025,Yee2025,Tamburo2025,Zhang2025,Wang2026}. However, it remains unclear whether these planets arise from a single universal formation and dynamical pathway. We are addressing this by constructing a statistically robust sample of small planets spanning diverse environmental conditions (e.g., stellar binarity and stellar effective temperature), allowing us to map the obliquity distribution and identify its dependence on system environment.

\subsection{Target Systems}

\begin{deluxetable*}{lccccccccl}
\tablecaption{Literature parameters for systems in our sample. Uncertainties on $P_\mathrm{orb}$ and $R_p$ are not reported, as we revise them in a later section. 
Reference Key: \textbf{A}---\cite{Hill2024}, \textbf{B}---\cite{MacDougall2023}, 
\textbf{C}---\cite{Polanski2024}, \textbf{D}---This work.\label{tab:literatureparams}}
\tablewidth{0pt}

\tablehead{
\colhead{Name} &
\colhead{Host $V$} &
\colhead{Host $T_\mathrm{eff}$} &
\colhead{$\rho_\star$} &
\colhead{$N_\mathrm{pl}$} &
\colhead{$P_\mathrm{orb}$} &
\colhead{$M_p$} &
\colhead{$R_p$} &
\colhead{$e$} &
\colhead{Reference(s)} \\ [-1.5ex]
\colhead{} &
\colhead{(mag)} &
\colhead{(K)} &
\colhead{(g cm$^{-3}$)} &
\colhead{} &
\colhead{(d)} &
\colhead{(M$_\oplus$)} &
\colhead{(R$_\oplus$)} &
\colhead{} &
\colhead{}
}

\startdata
TOI-1386 b & 10.6 & 5793\,$\pm\,75$ & 1.35\,$\pm$\,0.09 & 2 & 25.8 & 47.0\,$\pm$\,6.0 & 6.1 & 0.06\,$\pm$\,0.05      & A,B \\
TOI-480 b  & 7.3  & 6174\,$\pm$\,105 & 0.54\,$\pm$\,0.05 & 1 & 6.9  & 20.8\,$\pm$\,4.4 & 2.9 & $\equiv 0$      & B,C \\
TOI-4596 b & 9.8  & 5881\,$\pm$\,70 & 1.84\,$\pm$\,0.07  & 1 & 4.1  & --            & 2.9 & $\equiv 0$      & D \\
TOI-1823 b & 10.7 & 4927\,$\pm$\,58 & 2.24\,$\pm$\,0.30  & 1 & 38.8 & 67.4\,$\pm$\,8.3 & 7.5 & $\equiv 0$ & B,C \\
\enddata


\end{deluxetable*}

A brief summary of the SLOPE targets presented in this work is given in Table \ref{tab:literatureparams}, while the full description is given after our analysis. In this manuscript we present the first four observations of the obliquity survey for the stars TOI-1386, TOI-480, TOI-4596, and TOI-1823, which host planets ranging from sub-Saturn to sub-Neptune size. In total, the SLOPE survey aims to increase the number of measurements for this class of planets by $\sim$20 under a uniform analysis pipeline (described in Section \ref{sec:analysis}). We intend to also make the pipeline publicly available as a package for joint modeling photometry and RVs with the classical RM effect.

\subsection{Spectral Characterization with Keck-HIRES} \label{sec:hires}

Models of transiting planets can benefit substantially from independent constraints on intrinsic stellar properties  (e.g., \citealt{Petigura2020}, \citealt{Gilbert2022}, \citealt{MacDougall2023b}). In this section, we introduce new constraints that facilitate modeling in the sections that follow.

\subsubsection{Characterization of TOI-4596}

To characterize the host TOI-4596, we observed the star with the High Resolution Echelle Spectrometer\footnote{Characterization was performed on the HIRES spectra rather than KPF spectra because a \textsc{specmatch} implementation does not exist for the newer instrument.} (HIRES) at Keck Observatory \citep{Vogt1994} without the iodine cell on UT 18 January 2026. The entrance slit used was the C1 decker, which produced a spectral resolution of R$\sim$45,000. Over the wavelength regime of interest ($\sim$5000-6200 $\mathring{\text{A}}$) the spectrum had SNR $\gtrsim70$ per reduced pixel. To be consistent with previous analysis of the other host stars, we largely followed the stellar characterization methodology of the TKS survey \citep{MacDougall2023,Chontos2022}. In short, the stellar effective temperature $T_\text{eff}$, surface gravity $\log g$, and metallicity [Fe/H] were estimated using \textsc{specmatch-syn} \citep{Petigura2015,Petigura2017}, a package that matches the input spectrum to the high-resolution library of synthetic spectra from \cite{Coelho2005} via $\chi^2$ minimization. By modeling the spectral broadening and estimating the HIRES line-spread function, it also produces an estimate of the projected stellar rotational velocity \vsini\ with an uncertainty of $\sim$ 1 km~s$^{-1}$.

The best fit parameters and their uncertainties were then used as prior distributions for the isochrone fitting package \textsc{isoclassify} \citep{Huber2017,Berger2020,Berger2023}. As additional inputs to the grid, we included the Gaia DR3 parallax and $G$, $B_p$, and $R_p$ magnitudes \citep{GaiaDR3}, as well as the $J$, $H$, and $K$ band magnitudes from the Two-Micron All Sky Survey \citep{Skrutskie2006}. To account for systematic uncertainties, we set a minimum error of 0.01 on the Gaia magnitudes \citep{MacDougall2023}. We chose the PARSEC \citep{Bressan2012} isochrone grid to do the underlying interpolation. We experimented with the dust map from \cite{Green2019} using the \textsc{mwdust} package \citep{Bovy2016} and found that the results were consistent with negligible extinction. We ultimately allowed the extinction to be estimated internally using the spectral energy distribution (SED) informed by the photometry. \textsc{isoclassify} refines the spectroscopic parameters and outputs new posteriors for the stellar radius $R_\star$, stellar mass $M_\star$, luminosity $L_\star$, and age. The new stellar parameters are given in Table \ref{tab:toi4596_params}. In agreement with the estimate from the Transiting Exoplanet Survey Satellite (TESS) catalog \citep{Jenkins2016}, we find TOI-4596 to be closely solar-like in nature. We also find weak evidence that the star is somewhat young ($\sim$500 Myr).

\begin{deluxetable}{llr}
\tablecaption{Stellar parameters for TOI-4596 derived from the HIRES spectrum and photometry. All posteriors are those from the \textsc{isoclassify} package with the exception of \vsini\ which comes directly from \textsc{specmatch-syn}.}
\label{tab:toi4596_params}
\tablehead{
\colhead{Parameter} & \colhead{Unit} & \colhead{Value}
}
\startdata
$T_{\mathrm{eff}}$ & K & $5881^{+67}_{-73}$  \\
$\log g$ & log$_{10}$ (cm s$^{-2}$)          & $4.517^{+0.009}_{-0.011}$ \\
$\mathrm{[Fe/H]}$ & dex  & $-0.058^{+0.092}_{-0.056}$ \\
\vsini & km~s$^{-1}$           & $2.5\pm1.0$               \\
$R_\star$ & R$_\odot$          & $0.921^{+0.014}_{-0.014}$ \\
$M_\star$ &  M$_\odot$       & $1.018^{+0.021}_{-0.019}$ \\
$\rho_\star$ & g cm$^{-3}$       & $1.835^{+0.067}_{-0.068}$ \\
$L_\star$ & L$_\odot$         & $0.909^{+0.055}_{-0.050}$ \\
Age & Myr               & $483^{+663}_{-298}$ \\
Distance & pc          & $92.85^{+0.14}_{-0.14}$ \\
\enddata
\end{deluxetable}

\cite{Vach2024} identified TOI-4596 as a possible member of the young (55~Myr) cluster Theia 133.  However, this conclusion is disfavored after spectroscopic followup revealed no lithium features in the stellar spectrum. Due to the relatively low amplitude of photometric variability in the TESS data (up to $\sim$0.1\% variability in 2-minute photometry) and somewhat low \vsini, we come to a similar conclusion that the star is unlikely to be a member. Since a precise age is not the focus of our analysis, we simply comment on it without pursuing additional validation.


\subsubsection{Spectroscopic Constraints from the TKS Collaboration} \label{sec:lowimpactparam}

RM observations of transits with low impact parameters ($b \approx 0$) can have strongly degenerate values of $\lambda$ and \vsini, leading to a large tail in the \vsini\ posterior. This was a concern for both TOI-1386~b and TOI-1823~b, which transit at $b\sim0.1$ (see Section \ref{sec:results}). If spectroscopic data constrains the \vsini\ a priori, the posterior of $\lambda$ will suffer less strongly from that degeneracy. As part of the TKS survey, all other targets were previously analyzed using a \textsc{specmatch-syn} fit to a similar HIRES reconnaissance spectrum with a nearly identical analysis methodology. While the \vsini\ measurements were not reported in the survey manuscripts, \cite{MacDougall2023} and \cite{Polanski2024} determined an upper bound of \vsini\ $\lesssim$ 2 km~s$^{-1}$ for TOI-1386 and an upper bound of \vsini\ $\lesssim$ 2 km~s$^{-1}$ for TOI-1823. Those results indicate that rotational broadening has a small contribution to both stellar line profiles compared to the HIRES instrumental broadening. A long-tailed \vsini\ posterior is thus inconsistent with the spectrum---we use the \textsc{specmatch-syn} results as priors in Section \ref{sec:model} to establish an upper limit. In contrast, \cite{MacDougall2023} and \cite{Polanski2024} found TOI-480 to have a significant rotational constribution, estimating \vsini\ = 8.6$\pm$1.0 km s$^{-1}$. For that target we adopt an uninformative prior, allowing the RM data to constrain the rotational velocity.

\section{Analysis Pipeline} \label{sec:analysis}

\subsection{TESS Photometry} \label{sec:phot}

When the amplitude of the RM effect is small, a precise orbital solution is critical. For each target, we queried the 2-minute cadence photometry from TESS \citep{Ricker2015} using the \textsc{lightkurve} package \citep{lightkurve}. We chose the Presearch Data Conditioning Simple Aperture Photometry (PDCSAP) values from the Science Processing Operations Center (SPOC) pipeline \citep{Jenkins2016} as our flux measurements of choice, which are automatically detrended using co-trending basis vectors (CBVs) which reduces the impact of systematic trends. To facilitate precise transit modeling, we also fit cubic splines to the hourly binned out-of-transit data in each Sector (greater than a transit duration away from the transit times estimated using the literature ephemerids in Table \ref{tab:literatureparams}) to flatten the light curves before modeling. 

For all our targets, we detected high SNR transits in the 2-minute cadence data over long baselines. TOI-1386 was observed at 2-minute cadence in TESS Sectors 56, 57, 76, 77, 83, and 84---a baseline of 785 days (30 times the orbital period). TOI-480 was observed in Sectors 6, 33, and 87---a baseline of 2222 days (324 times the orbital period). TOI-4596 was observed in Sectors 43, 71, and 72---a baseline of 811 days (197 orbital periods). TOI-1823 was observed in TESS Sectors 14, 15, 21, 22, 41, 48, 49, 75, and 76---a total baseline of 1711 days (44 times the orbital period). These baselines are sufficiently long to constrain the orbital solutions of the planets. We tested for the presence of transit timing variations (TTV) but did not see evidence (see Appendix \ref{sec:ttvs}).

We also attempted to constrain the rotational period of our targets using periodic modulations in the TESS photometry. Our criterion for a detection was that 1) a single peak in the Lomb-Scargle periodogram \citep{Zechmeister2009} of the PDCSAP time series dominates the power spectrum, 2) when phase folded at the trial period, a coherent sinusoid was visible and higher harmonics did not produce a coherent signature to avoid aliasing, and 3) when fitting a sinusoid to the data, the amplitude of that sinusoid is both 5$\sigma$ from zero and large compared to the TESS measurement uncertainties. We refrain from a detailed discussion of our detection thresholds because none of our targets demonstrated significant candidate periods. Thus, the stellar inclinations remain unknown.

\subsection{KPF Radial Velocities}

Using the orbital solutions from the references in Table \ref{tab:literatureparams}, we observed each of our four targets during a planetary transit with the KPF spectrograph at W. M. Keck Observatory \citep{Gibson2024}. Whenever possible, we also collected observations of the star before and/or after the transit event to constrain that night's RV zero-point and to monitor trends induced by instrumental drift over the several hours of in-transit observations. KPF collects spectra in two channels, green (450--600 nm) and red (600--870 nm). The octagonal fiber that is illuminated by the target star (hereafter, the science fiber) is also split into three `traces', allowing three copies of the stellar spectrum across the each of the green and red channels (R$\sim$95,000) during each observation. 

Generally, the KPF Data Reduction Pipeline\footnote{\url{https://github.com/Keck-DataReductionPipelines/KPF-Pipeline}} (DRP) computes wavelength solutions twice a day using spectra of a Laser Frequency Comb (LFC) illuminating all the science traces. The LFC is an extremely accurate calibrator, but cannot be used at higher cadence due to its finite lifetime. This limits how precisely the wavelength solution can be corrected to account for drifting of the instrument RV zero-point throughout the night. To mitigate that effect, we collected calibration exposures with the KPF Fabry-Perot etalon lamp at either end of each transit sequence. In these exposures, the etalon source illuminates the science traces for instantaneous, precise wavelength calibration. Critically, these calibrations are closer in time to our observations than the standard LFC exposures, meaning they capture more relevant thermal behavior in the spectrometer. 

We then modeled the instrumental drift over the course of the transit in each channel/trace combination. Using the drift computed from the etalon frames (measured as changes in the line centers in pixel space), we adjusted the wavelength solution of the reduced one-dimensional spectra accordingly. A global implementation of this technique is in development for the public version of the DRP, which will incorporate the drift calibrations over monthly timescales (Handley et al., in prep.). After adjusting the wavelength solution, the one-dimensional spectra were again run through the final stages of the KPF pipeline to produce RVs which are less strongly affected by instrumental drifts. In brief, RVs were computed via cross-correlation with an ESPRESSO \citep{Pepe2021} stellar mask of the appropriate spectral type, given the effective temperature in Table \ref{tab:literatureparams}.

\begin{deluxetable*} {lcccc}
\tablecaption{KPF RV measurements.}
\label{tab:kpfdata}
\tablehead{
\colhead{Star} & \colhead{Time (BJD)} & \colhead{RV (m~s$^{-1}$)} & \colhead{RV Error (m~s$^{-1}$)} & \colhead{Exposure Time (min)}
}
\startdata
TOI-1386 & 2460586.70404 & -23605.3 & 2.04 & 5\\
TOI-1386 & 2460586.70808 & -23608.34 & 1.92 & 5\\
TOI-1386 & 2460586.71252 & -23606.7 & 2.03 & 5\\
TOI-1386 & 2460586.71655 & -23608.05 & 1.98 & 5\\
TOI-1386 & 2460586.72059 & -23608.58 & 1.9 & 5\\
TOI-1386 & 2460586.72462 & -23606.55 & 1.87 & 5\\
TOI-1386 & 2460586.72866 & -23607.63 & 1.97 & 5\\
TOI-1386 & 2460586.73269 & -23606.63 & 1.99 & 5\\
TOI-1386 & 2460586.73672 & -23604.47 & 1.91 & 5\\
$\cdots$   & $\cdots$ & $\cdots$ & $\cdots$ & $\cdots$ \\
\enddata
\tablecomments{The full table is available in machine-readable form.}
\end{deluxetable*}

The duration of the exposures was chosen so that the Poisson noise per reduced spectral pixel would be significantly above the read noiseand a large number of exposures ($>10$) could be taken during transit to get fine temporal resolution throughout. We favored these criterion above precise pressure mode averaging \citep{Chaplin2019} because the amplitudes of those oscillations would be small ($\lesssim 50$cm s$^{-1}$) compared to our photon error budget. The full time series of the KPF RVs for all targets is given in Table \ref{tab:kpfdata}. A concise summary of the observational campaigns is as follows. TOI-1386 was observed on UT 03 October 2024 with 300 s exposures over a 7.6-hour baseline, including 1.9 hr out of transit. A total of 79 spectra were obtained, yielding trace-combined signal-to-noise ratios (SNRs) per reduced pixel of $\sim$120 (green) and $\sim$140 (red) at the blaze peak of each spectral order. On UT 08 December 2024, TOI-480 was monitored for 6.6 hr with 180 s integrations, including 3.1 hr out of transit. This resulted in 102 exposures, achieving high SNRs of approximately 350 in the green and 450 in the red channels. On UT 19 December 2025, TOI-4596 was observed over a 4.5-hour sequence with 180 s integrations, of which 2.3 hr occurred outside transit. This campaign produced 70 spectra with SNRs of $\sim$110 (green) and $\sim$140 (red). On UT 22 April 2024, TOI-1823 was observed continuously for 8.5 hr using 600 s integrations, of which 2.5 hr occurred outside of transit. This sequence produced 43 spectra with SNRs of approximately 150 in the green channel and 210 in the red.

\subsection{Joint Modeling} \label{sec:model}

We analyzed the SLOPE targets consistently with a pipeline that models the detrended TESS photometry and KPF RVs jointly. The pipeline is designed to be a simple and intuitive tool that will be used for future survey targets as well as made publicly available. This tool samples the transit parameters $\theta_\text{T}$ in the basis

\begin{equation} \label{eq:transitbasis}
    \theta_\text{T} = \{T_0,P_\text{orb},R_\text{p}/R_\star,a/R_\star,b,q_{1},q_{2}\},
\end{equation}
where $T_0$ is the transit midpoint, $P_\text{orb}$ is the orbital period, $R_\text{p}/R_\star$ is the planet-to-stellar radius ratio, $a/R_\star$ is the orbital-to-stellar radius ratio, $b$ is the transit impact parameter, and $q_1$ and $q_2$ parametrize the quadratic limb-darkening model in \cite{Kipping2013}. For each choice of $\theta_\text{T}$, the pipeline generated a theoretical light curve using the \textsc{batman} package \citep{batman}. Although our tool does allow for nonzero orbital eccentricity, we did not include eccentricity as a free parameter. As shown in Table \ref{tab:literatureparams}, our targets have orbits consistent with zero eccentricity, so we adopted circular orbits. An initial transit model of TOI-4596~b also indicated that an eccentric orbit was not needed to explain the TESS observations, so we assumed it to be circular as well.

\begin{figure*} \label{fig:transitfits}
    \centering
    \includegraphics[width=\linewidth]{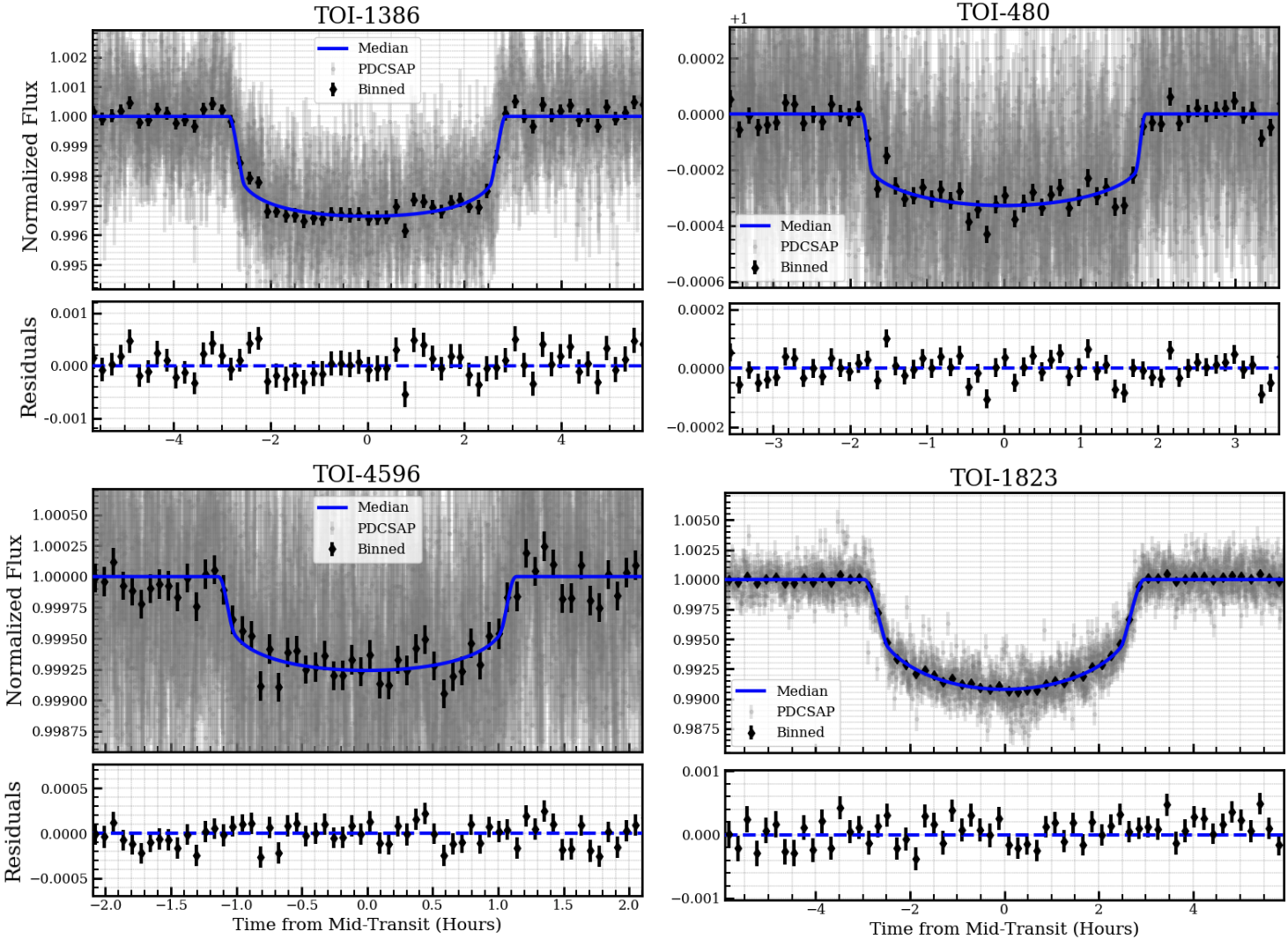}
    \caption{Our spline-corrected TESS light curves, phase-folded at the median period from the posterior distributions of the joint model (no targets showed evidence of TTVs, see Appendix \ref{sec:ttvs}). The median transit model is plotted in blue. A binned dataset is also shown (bins were calculated from the phase-folded data at 1/20 of the transit duration).}
\end{figure*}

For the RM model, the pipeline used the prescription from \cite{Hirano2011}, which includes the effects of rotational broadening, macroturbulence, and the instrumental profile of KPF. Our pipeline estimates the macroturbulent velocity from the effective temperature following the scaling in \cite{Valenti2005}, sets the microturbulent velocity to 0.7 km s$^{-1}$, and assumes an intrinsic line width of 1 km s$^{-1}$. These choices, while reasonable, are not particularly consequential. Large changes in the values of the profile parameters modulate the RM signal at the $\sim$few \% level (here, $\lesssim$ 10 cm s$^{-1}$; \cite{Hirano2011}), which we would not be sensitive to. However, we do include the effects of convective blueshift using the model from \cite{Shporer2011}. An offset and trend term for the RV time series are also included, which allow persisting instrumental artifacts and Keplerian motion of the host to be modeled.

We noted that the DRP-reported RV uncertainties were overestimated in the early stages of fitting the observations (see, e.g., \citealt{Handley2025,Giacalone2025,Yee2025}), and thus opted to estimate them such that the reduced $\chi^2$ statistic was roughly 1. The SNR was roughly constant throughout each dataset, so we assume the RV uncertainty to also be constant for each target. To do this, we used a likelihood function in which the RV uncertainty for each star $\sigma_\mathrm{RV}$ is a free parameter.  That is, our RM fitting vector $\theta_\text{RM}$ was

\begin{equation} \label{eq:rmbasis}
    \theta_\text{RM} = \theta_\text{T}\cup\{\vsinimath,\lambda,v_\text{cb},\gamma,\dot{\gamma},\sigma_\text{RV},q_1',q_2'\,\},
\end{equation}
where \vsini\ is the projected rotational velocity of the star, $\lambda$ is the sky-projected stellar obliquity angle, $v_\text{cb}$ is the magnitude of the convective blueshift velocity, $\gamma$ is an RV offset term, and $\dot{\gamma}$ is an apparent linear component\footnote{In the observations of TOI-4596 we observed a clearly non-linear trend in the out-of-transit RVs, so for that target only, we include a quadratic term $\ddot{\gamma}$ as well.} of the RV time series induced by the instrument. Limb darkening parameters $q_1'$ and $q_2'$ refer to those as estimated in the KPF RV bandpass (subtly different from the photometric $q_1$ and $q_2$). To improve convergence, we sampled $\lambda$ and \vsini\ using the parametrization $\{$\vsincos,\vsinsin$\}$. With these definitions, the likelihood is
\begin{equation} \label{eq:likelihood}
    \begin{split}
    -2 \,\text{ln}&\, \mathcal{L}(\theta_\text{RM}) =  \sum_{j}\left( \frac{\text{F}_\text{Phot}(t_j) - h(t_j;\theta_\text{RM})}{\sigma_{\text{Phot},j}} \right)^2 \\
    &+\sum_{i}\left[ \left( \frac{\text{RV}(t_i) - g(t_i;\theta_\text{RM})}{\sigma_\mathrm{RV}} \right)^2 + \text{ln}\, 2\pi\sigma_\mathrm{RV}\right],
    \end{split}
\end{equation}
where, in our example, F$_\text{Phot}$($t_j$) is the PDCSAP flux measured by TESS (after our spline correction) at epoch $t_j$, $h$($t_j,\theta_\text{RM}$) is the transit light curve model evaluated at that time, $\sigma_{\text{Phot},j}$ is the flux uncertainty reported by TESS at each of those epochs, RV($t_i$) is the RV measured with KPF at a time $t_i$, and $g$($t_i,\theta_\text{RM}$) is the theoretical RM model evaluated at those times. The model $g$($t_i,\theta_\text{RM}$) was also oversampled to a cadence of roughly one minute and averaged over the exposure time, replicating the impact of the RV evolving over a single observation.

For each target, we assembled a starting vector $\theta_\text{T}$ using the median transit parameters from the references in Table \ref{tab:literatureparams}. We initialized the limb darkening parameters $q_1$ and $q_2$ using the theoretically expected values given by the tabulations in \cite{Exofast} as functions of the stellar properties. For the transit parameters, we use uninformative uniform priors. The one exception is that we convert the parameter $a/R_\star$ into the parameter $q_\star$ (the stellar density; see Section \ref{sec:hires}), for which we use Gaussian priors based on the stellar parameters, which are independent of the TESS or KPF time series. All priors are listed in Table \ref{tab:priors}. 

As for the remainder of $\theta_\text{RM}$, \vsini\ was initialized at 2 km s$^{-1}$ and given a uniform prior with the upper bound informed by the \textsc{specmatch-syn} results in Section \ref{sec:hires}; $\lambda$ was initialized at 0$\degree$, and allowed to span the full extent of $\{-180\degree,180\degree\}$; $v_\text{cb}$ was given a Gaussian prior of width 100~m~s$^{-1}$ centered at an estimated velocity from the empirical study in \cite{Liebing2021} as informed by the host $T_\text{eff}$; $\gamma$ and $\dot{\gamma}$ were given no prior, and initialized at the median of the KPF RV time series and zero, respectively. Using the \textsc{scipy.optimize} package, we maximized the sum of the log-likelihood and log-prior to get the maximum-a-posteriori (MAP) value of $\theta_\text{RM}$.

Using the MAP solution as the starting vector, we estimated the properties of the posterior distributions under a Markov Chain Monte Carlo (MCMC) framework with \textsc{emcee} as the backend \citep{foremanmackey2013}. We utilized 48 independent walkers running for 10,000 steps each with the first 5,000 discarded as burn-in and the remaining steps thinned by a factor of 5. We verified convergence of the MCMC chains using the Gelman-Rubin statistic ($\hat{R}<1.01$ for all parameters in $\theta_\text{RM}$ for all systems).

\section{Results} \label{sec:results}

\begin{figure*} \label{fig:rmfits}
    \centering
    \includegraphics[width=\linewidth]{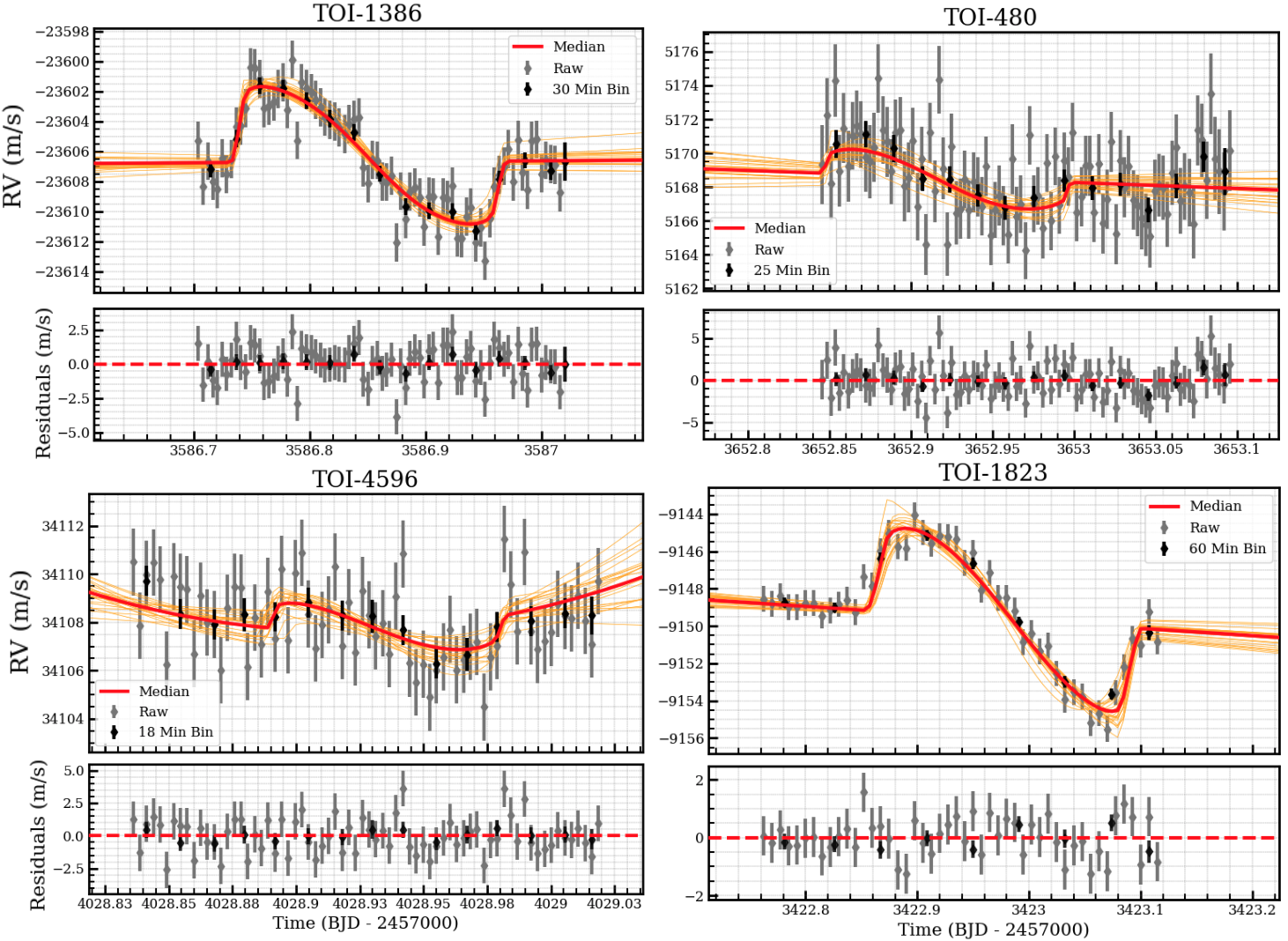}
    \caption{KPF RV time-series during each transit event. Red lines indicate the RM model of the transiting planet using the median of the posterior distributions of the joint model, while the orange lines indicate random draws from those distributions. Although the model is fit to the raw RVs, binned data points (black) are also included for clarity. The curved out-of-transit baseline during the observations of TOI-4596 (bottom left) lead us to choose a quadratic model for that system alone.}
\end{figure*}

Our pipeline confidently detected the RM effect in the KPF observations of the four systems. Plots of the RVs during each transit sequence along with the best fit models are in Figure~\ref{fig:rmfits}, while the phase-folded TESS photometry is plotted in Figure~\ref{fig:transitfits}. A full summary of the posterior distributions for the transit and RM parameters can be found in Table \ref{tab:posteriors}, and corner plots of the distributions along with their covariances can be found in the Appendix. We find that all four planets induce an RM signal which is consistent with an aligned orbit at $<2\sigma$. Furthermore, the pipeline recovered our input prior on $v_\text{cb}$ with no visible covariance with $\lambda$ for all our targets, indicating that there is no strong evidence of contamination of the RM signal due to convective blueshift. In the following sections, we discuss each measurement in context.

\begin{table*}
\centering
\caption{Joint model priors adopted for all targets. $\mathcal{U}$ indicates a uniform prior between two values, while $\mathcal{G}$ indicates a Gaussian prior parametrized by a mean and variance respectively. The scaled semi-major axis parameter $a/R_\star$ was sampled in the MCMC chains, but a prior was set on a $\rho_\star$ reparameterization to leverage knowledge of the host star (values of $R_\star$ and $M_\star$ in Table~\ref{tab:literatureparams}). Limb darkening coefficients are relabeled according to the instrument for clarity. The magnitude of the convective blueshift velocity $v_\text{cb}$ was not allowed to pass below zero in addition to the listed prior.}
\label{tab:priors}
\begin{tabular}{lcccc}
\toprule
\multicolumn{5}{c}{\textbf{Priors}} \\
Parameter & TOI-1386 b & TOI-480 b & TOI-4596 b & TOI-1823 b \\
\midrule
$\lambda\,(\mathrm{deg})$ & $\mathcal{U}(-\pi,\pi)$ & $\mathcal{U}(-\pi,\pi)$ & $\mathcal{U}(-\pi,\pi)$ & $\mathcal{U}(-\pi,\pi)$ \\
$v\sin i_\star\,(\mathrm{km\,s^{-1}})$ & $\mathcal{U}(0,4)$ & $\mathcal{U}(0,15)$ & $\mathcal{U}(0,15)$ & $\mathcal{U}(0,4)$ \\
$\gamma\,(\mathrm{m\,s^{-1}})$ & -- & -- & -- & -- \\
$\dot{\gamma}\,(\mathrm{m\,s^{-1}\,d^{-1}})$ & -- & -- & -- & -- \\
$\ddot{\gamma}\,(\mathrm{m\,s^{-1}\,d^{-2}})$ & $\equiv0$ & $\equiv0$ & -- & $\equiv0$ \\
$\sigma_{\mathrm{KPF}}\,(\mathrm{m\,s^{-1}})$ & -- & -- & -- & -- \\
$v_{\mathrm{cb}}\,(\mathrm{m\,s^{-1}})$ & $\mathcal{G}(317,100^2)$ & $\mathcal{G}(641,100^2)$ & $\mathcal{G}(249,100^2)$ & $\mathcal{G}(93,100^2)$ \\
$b$ & $\mathcal{U}(0,1)$ & $\mathcal{U}(0,1)$ & $\mathcal{U}(0,1)$ & $\mathcal{U}(0,1)$ \\
$\rho_\star\,(\mathrm{g\,cm^{-3}})$ & $\mathcal{G}(1.35,0.09^2)$ & $\mathcal{G}(0.54,0.05^2)$ & $\mathcal{G}(1.84,0.07^2)$ & $\mathcal{G}(2.24,0.30^2)$ \\
$R_p/R_\star$ & $\mathcal{U}(0.001, 0.2)$ & $\mathcal{U}(0.001, 0.2)$ & $\mathcal{U}(0.001, 0.2)$ & $\mathcal{U}(0.001, 0.2)$ \\
$q_{1,\mathrm{KPF}}$ & $\mathcal{U}(0,1)$ & $\mathcal{U}(0,1)$ & $\mathcal{U}(0,1)$ & $\mathcal{U}(0,1)$ \\
$q_{2,\mathrm{KPF}}$ & $\mathcal{U}(0,1)$ & $\mathcal{U}(0,1)$ & $\mathcal{U}(0,1)$ & $\mathcal{U}(0,1)$ \\
$q_{1,\mathrm{TESS}}$ & $\mathcal{U}(0,1)$ & $\mathcal{U}(0,1)$ & $\mathcal{U}(0,1)$ & $\mathcal{U}(0,1)$ \\
$q_{2,\mathrm{TESS}}$ & $\mathcal{U}(0,1)$ & $\mathcal{U}(0,1)$ & $\mathcal{U}(0,1)$ & $\mathcal{U}(0,1)$ \\
$P_{\mathrm{orb}}\,(\mathrm{d})$ & $\mathcal{U}(25.828 , 25.848)$ & $\mathcal{U}(6.856 , 6.876)$ & $\mathcal{U}(4.109 , 4.129)$ & $\mathcal{U}(38.803 , 38.823)$ \\
$T_0\,(\mathrm{BJD}-2457000)$ & $\mathcal{U}(1751.82 , 1752.82)$ & $\mathcal{U}(1469.07 , 1470.07)$ & $\mathcal{U}(3282.82 , 3283.82)$ & $\mathcal{U}(1714.70 , 1715.70)$ \\
\bottomrule
\end{tabular}
\end{table*}

\begin{table*}
\centering
\caption{Posterior constraints for all targets from our joint model, reported as medians with 68\% credible intervals. Most of our marginalized posteriors are well approximated as Gaussian, but covariances are sometimes present (see Appendix).}
\label{tab:posteriors}
\begin{tabular}{lcccc}
\toprule
\multicolumn{5}{c}{\textbf{Posteriors}} \\
Parameter & TOI-1386 b & TOI-480 b & TOI-4596 b & TOI-1823 b \\
\midrule
$\lambda\,(\mathrm{deg})$ 
& $23.0^{+13.9}_{-14.8}$ 
& $-6.6^{+12.9}_{-13.0}$ 
& $-3.4^{+20.1}_{-20.8}$ 
& $8.4^{+22.2}_{-14.3}$ \\

$v\sin i_\star\,(\mathrm{km\,s^{-1}})$ 
& $2.38^{+0.29}_{-0.21}$ 
& $8.79^{+1.06}_{-1.10}$ 
& $3.43^{+0.87}_{-0.90}$ 
& $0.90^{+0.14}_{-0.05}$ \\

$\gamma\,(\mathrm{m\,s^{-1}})$ 
& $-23606.8^{+0.4}_{-0.5}$ 
& $5169.1^{+0.7}_{-0.7}$ 
& $34109.3^{+0.6}_{-0.6}$ 
& $-9148.6^{+0.2}_{-0.2}$ \\

$\dot{\gamma}\,(\mathrm{m\,s^{-1}\,d^{-1}})$ 
& $0.4^{+2.1}_{-2.1}$ 
& $-3.6^{+3.5}_{-3.6}$ 
& $-32.4^{+20.3}_{-20.9}$ 
& $-3.9^{+1.4}_{-1.4}$ \\

$\ddot{\gamma}\,(\mathrm{m\,s^{-1}\,d^{-2}})$ 
& $\equiv0$ 
& $\equiv0$
& $162^{+109}_{-106}$
& $\equiv0$ \\

$\sigma_{\mathrm{KPF}}\,(\mathrm{m\,s^{-1}})$ 
& $1.27^{+0.11}_{-0.10}$ 
& $1.99^{+0.16}_{-0.14}$ 
& $1.38^{+0.13}_{-0.12}$ 
& $0.69^{+0.08}_{-0.07}$ \\

$v_{\mathrm{cb}}\,(\mathrm{m\,s^{-1}})$ 
& $353^{+95}_{-100}$ 
& $637^{+99}_{-97}$ 
& $238^{+99}_{-97}$ 
& $133^{+67}_{-72}$ \\

$b$ 
& $0.15^{+0.08}_{-0.08}$ 
& $0.67^{+0.03}_{-0.03}$ 
& $0.66^{+0.06}_{-0.07}$ 
& $0.13^{+0.16}_{-0.11}$ \\

$a/R_\star$ 
& $36.23^{+0.42}_{-0.52}$ 
& $11.04^{+0.34}_{-0.32}$ 
& $10.99^{+0.72}_{-0.83}$ 
& $54.07^{+0.64}_{-1.55}$ \\

$R_p/R_\star$ 
& $0.0539^{+0.0006}_{-0.0006}$ 
& $0.0176^{+0.0003}_{-0.0003}$ 
& $0.0266^{+0.0008}_{-0.0007}$ 
& $0.0853^{+0.0011}_{-0.0007}$ \\

$q_{1,\mathrm{KPF}}$ 
& $0.44^{+0.27}_{-0.19}$ 
& $0.51^{+0.34}_{-0.33}$ 
& $0.53^{+0.31}_{-0.35}$ 
& $0.46^{+0.33}_{-0.31}$ \\

$q_{2,\mathrm{KPF}}$ 
& $0.47^{+0.34}_{-0.29}$ 
& $0.49^{+0.34}_{-0.34}$ 
& $0.50^{+0.34}_{-0.35}$ 
& $0.48^{+0.36}_{-0.34}$ \\

$q_{1,\mathrm{TESS}}$ 
& $0.27^{+0.19}_{-0.12}$ 
& $0.31^{+0.14}_{-0.10}$ 
& $0.44^{+0.28}_{-0.21}$ 
& $0.40^{+0.12}_{-0.09}$ \\

$q_{2,\mathrm{TESS}}$ 
& $0.30^{+0.30}_{-0.19}$ 
& $0.49^{+0.32}_{-0.32}$ 
& $0.29^{+0.35}_{-0.20}$ 
& $0.52^{+0.14}_{-0.11}$ \\

$P_{\mathrm{orb}}\,(\mathrm{d})$ 
& $25.83843^{+0.00005}_{-0.00005}$ 
& $6.865903^{+0.000005}_{-0.000005}$ 
& $4.119409^{+0.000006}_{-0.000007}$ 
& $38.81360^{+0.00002}_{-0.00002}$ \\

$T_0\,(\mathrm{BJD}-2457000)$ 
& $1752.322^{+0.003}_{-0.003}$ 
& $1469.5643^{+0.0007}_{-0.0007}$ 
& $3283.3223^{+0.0008}_{-0.0008}$ 
& $1715.1787^{+0.0004}_{-0.0004}$ \\
\bottomrule
\end{tabular}
\end{table*}

\subsection{TOI-1386~b}

TOI-1386 is a solar-type star ($T_\text{eff}\sim5800$ K, $V=10.6$) that hosts two known planets \citep{Hill2024}. The target planet TOI-1386~b is a warm Neptune with a mass of $M_p=47M_\oplus$ orbiting at $P_\text{orb}=25.8$ days in a nearly circular ($e\lesssim0.05$) configuration. Exterior to that planet is an Saturn-mass ($M_p=98M_\oplus$) planet with an eccentric ($e\sim0.3$; \citealt{Hill2024,Polanski2024}), 230-day orbit. Modeling the KPF RV time series, we find that TOI-1386~b orbits with a slightly non-zero value of $\lambda=23.0^{+13.9}_{-14.8}$ degrees. The uncertainty is driven primarily by the planet transiting with a low impact parameter ($b\sim\,$0.1).  The model also found \vsini= 2.38$^{+0.29}_{-0.21}$~km~s$^{-1}$, a 10$\,\sigma$ detection of the RM effect. If the planet is indeed misaligned, mutual interactions with TOI-1386~c might help to determine the dynamical origin.  Given that the inner planet has a nearly circular orbit yet is at wide separation, a quiescent history would be favored because a strongly perturbed orbit (e.g., scattering or ZKL; \citealt{Chatterjee2008,Fabrycky2007}) would not be strongly affected by tidal circularization \citep{Zahn1977,Hansen2010,Hansen2012,Ogilvie2014}. \cite{Hill2024} found that for coplanar geometry, the outer planet can drive significant eccentricity oscillations on the inner planet ($e_\text{max}\sim0.2$). Therefore, our observation of $\lambda$ could suggest a fully (both the star and planets) coplanar geometry being observed in a low eccentricity epoch for TOI-1386~b.

\subsection{TOI-480~b}

TOI-480 is a hot and bright F-type star ($T_\text{eff}\sim6200$ K, $V=7.3$; \citealt{MacDougall2023}) that hosts a single sub-Neptune planet at $P_\text{orb}=6.9$ days with a mass of $M_p\sim21M_\oplus$ \citep{Polanski2024}. Our RM measurement of $\lambda=6.6^{+12.9}_{-13.0}$ for TOI-480~b suggests that the planet's orbit is aligned with the stellar spin axis. TOI-480~b orbits a hot star, which commonly exhibit large misalignments from the orbits of short-period giant planets \citep{Winn2010b,Schlaufman2010}. Although evidence of trends for small planets above the Kraft break is slowly growing (e.g., \citealt{Dugan2025}), the number of measurements is not large enough to draw conclusions. Consistent with the expectation of fast rotation of the hot host star (these stars tend to have smaller convective envelopes and thus inefficient spin-down via magnetic braking; \citealt{Kraft1967,Barnes2003,Winn2015}), we measure \vsini =  8.79$^{+1.06}_{-1.10}$~km~s$^{-1}$, an $8\,\sigma$ detection. The rotational broadening is also the primary driver of the RV uncertainties; we found $\sigma_\text{KPF}\sim2$m~s$^{-1}$. 


\subsection{TOI-1823~b}

TOI-1823 is an early K-type star ($T_\text{eff}\sim4900$ K, $V=10.7$; \citealt{MacDougall2023}) which hosts a sub-Saturn planet ($R=7.5R_\oplus$, $M=67M_\oplus$; \citealt{Polanski2024}) on a relatively wide and circular 39 day orbit. We find TOI-1823~b to have a sky-projected obliquity of $\lambda=8.4^{+22.2}_{-14.3}$ degrees, consistent with perfect alignment. The observations were collected under excellent conditions (with an average seeing of 0.55"), and our model estimated an RV uncertainty of $\sigma_\text{KPF}\sim\,$0.65~m~s$^{-1}$. Similar to TOI-1386~b, the uncertainty in obliquity is primarily driven by the low transit impact parameter. We find \vsini\ = 0.90$^{+0.14}_{-0.05}$~km~s$^{-1}$ (over $10\,\sigma$ detection), and although our measurement strongly favors an aligned $\lambda$, the true obliquity will be between $\lambda$ and $90\degree$; the lower \vsini\ (the measured value implies a slow rotation period of 45 days) may indicate a misaligned stellar inclination and thus a more misaligned true obliquity. Distinguishing requires constraining the stellar rotation period, which was not resolved by the TESS photometry (Section \ref{sec:phot}).

\subsection{Confirmation of a Sub-Neptune Orbiting TOI-4596}

TOI-4596 is a young, sun-like star (Age $\sim$ 500\,Myr; $T_\text{eff}\sim5800$ K; $V=9.8$) that hosts a sub-Neptune ($R=2.9R_\oplus$) planet with a 4.1 day orbit. The candidate planet TOI-4596.01 was flagged by the Transiting Exoplanet Survey Satellite (TESS; \citealt{Jenkins2016,Ricker2015,Stassun2019}) Science Processing Operations Center (SPOC) pipeline. We detected the planet strongly in the KPF time series, and find the sub-Neptune is likely aligned with a posterior distribution of $\lambda=8.4^{+22.1}_{-14.3}$. The \vsini\ as measured by the RM effect is 3.43$^{+0.87}_{-0.90}$~km~s$^{-1}$, which is consistent with the spectral broadening estimate in Section \ref{sec:hires}. 

Because the planet was not previously validated, we vetted the possibility of the TESS transit signature being a false positive using the \textsc{triceratops} package \citep{Giacalone2021code}. \textsc{triceratops} models the target star and nearby potential contaminating stars, calculating the marginal likelihood of each transit-producing scenario. The code computes the False-Positive Probability (FPP; likelihood of a non-transiting planet origin from the target) and the Nearby False-Positive Probability (NFPP; likelihood of the brightness variation belonging to a contaminant star). We followed the recommended methodology, combining information from the light curves in Sectors 43, 71, and 72, and computed the FPP and NFPP for a large number of iterations to determine a statistical average. For TOI-4596~b, \textsc{triceratops} estimated a FPP = 0.027 and an NFPP $<10^{-6}$. \cite{Giacalone2021} suggests an FPP of $< 0.015$ and NFPP $< 10^{-3}$ for a planet to be validated via transits alone. Thus, while the risk of a false positive is low, the TESS data are not sufficient on their own. Our RM observations, which detected \vsini\ from the event at nearly 4$\,\sigma$, confirm the validity of the sub-Neptune.

\section{Ensemble Analysis} \label{sec:ensemble}

Here, we contextualize our obliquity measurements in the context of all measurements of small-planet obliquities. First, we statistically validate the claim that small, compact multi-planet systems are preferentially aligned. Then, we point out several emerging demographic trends that will be investigated in future SLOPE Survey studies.

\subsection{Alignment of Compact Multi-planet Systems} \label{sec:statisticalvalidation}

The persistence of compact planetary architectures over billions of years is a strong indicator of quiescent evolutionary pathways. Compact multi-planet systems are thought to form and migrate within their natal disks, maintaining coplanarity with one another and likely with the host star (although there are counterexamples, such as K2-290~A; \cite{Hjorth2021}). Therefore, the stellar obliquities of planets in compact multi-planet systems probe the relative alignment of stars and protoplanetary disks in the absence of major dynamical upheavals. Are compact systems of sub-Saturn to super-Earth planets aligned as one might expect? \cite{Radzom2024} found tentative evidence ($2.6\,\sigma$) among sub-Saturn planets when grouped by mass. \cite{Polanski2025} performed a similar test, achieving slightly higher significance by grouping the planets with radii $R_p\le4R_\oplus$. Recently, many more small planets have had their stellar obliquities measured, granting an enriched view of the demographics.

We constructed an updated sample, combining our SLOPE measurements with the TEPCAT catalog\footnote{Specifically, we downloaded the catalog on 02 February 2026.} of stellar obliquity measurements \citep{Southworth2011}, and cross-matched targets with the NASA Exoplanet Archive\footnote{\url{https://exoplanetarchive.ipac.caltech.edu/}} to get the best known masses and radii of the planets. We limited our sample to planets that are sub-Saturn-like and smaller, that is, a planetary mass $M_p < 95 M_\oplus$. For internal consistency, we considered only measurements of the sky-projected obliquities\footnote{Of course, an aligned value of $\lambda$ does not strictly prove the orbit is aligned in three dimensions, although it is likely. The true obliquity, $\psi$, is always closer to $90\degree$. A misaligned value of $\lambda$ means certain misalignment of $\psi$, but the opposite is not always true.} which were observed with the RM effect (avoiding strongly biased techniques, e.g., \citealt{Siegel2023}), and we neglected any systems where the error on $\lambda$ was greater than $40\degree$. To be consistent with previous works such as \cite{Radzom2024} and \cite{Wang2022}, we defined a compact system to be one in which the ratio of an outer planets orbital period to an inner planet $P_2/P_1$ is less than 4. In total, we considered 53 systems after our precision cuts, of which 20 systems were deemed compact and 33 were considered isolated.

The term `misalignment' can depend on the context. With this experiment, we hoped to distinguish systems that could plausibly have retained their obliquity at formation from those that were excited at a later epoch. This is challenging, as even planetary systems that evolved quiescently (such as the Solar System) can demonstrate stellar spin-orbit misalignments of $\sim$10$\degree$. Too strict a constraint on $|\lambda|$ leads to the potential misidentification of these systems. We define misalignment as $|\lambda|>0\degree$ with at least 2$\sigma$ confidence and $|\lambda|>20\degree$ with a minimum 1$\sigma$ confidence. That is, a `misaligned' system must be inclined enough to clearly bring a flat architecture into doubt.

Of the 20 planets in compact multi-planetary systems we considered, only one planet was considered misaligned, HD 3167~c \citep{Dalal2019,Bourrier2021}. That system remains difficult to interpret, as the reported 90$\degree$ mutual inclination between planets HD 3167~b and HD 3167~c has been questioned due to the improbability of the system being multi-transiting \citep{Teng2025,Zhang2025}. Further, the planets which orbit nearest HD~3167~c (for which we deemed the system `compact') are non-transiting \citep{Vanderburg2016,Christiansen2017,Bourrier2022}, indicating the planets may not necessarily be coplanar. Although it is indeed peculiar, it does not diminish the statistical result of our tests, so we kept it in the sample. In contrast, among the 19 aligned compact systems, all are consistent with $\lambda<10\degree$ at roughly 1$\sigma$, thus their status as aligned is probably not sensitive to our own definition of misalignment. 

\begin{figure*}
    \centering
    \includegraphics[width=\linewidth]{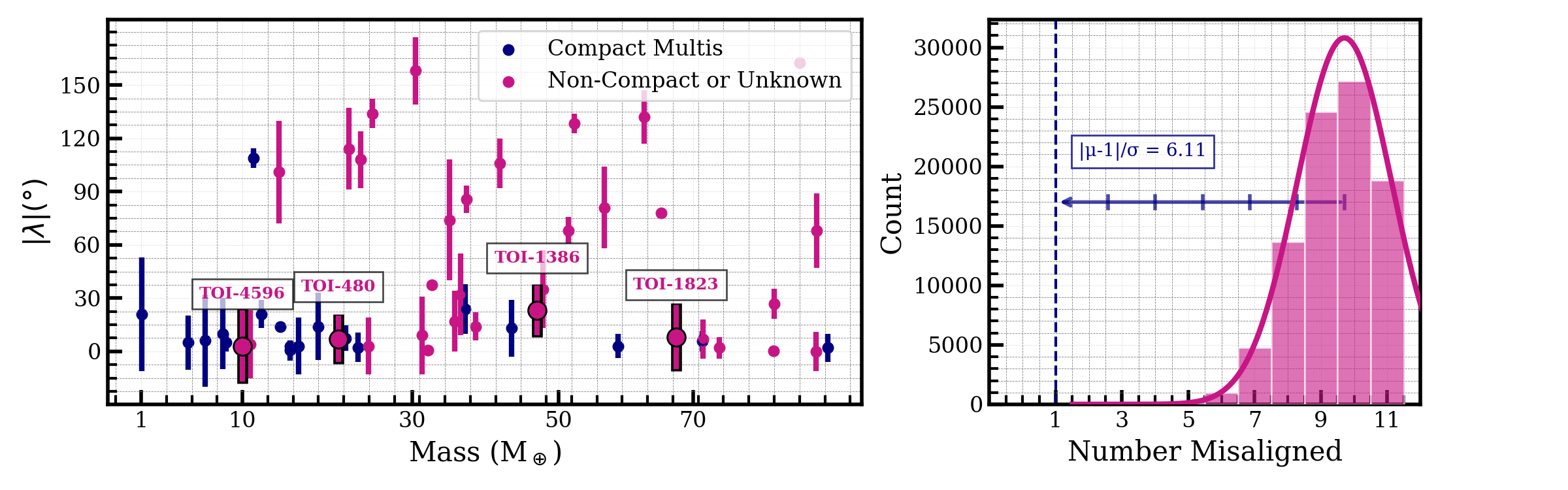}
    \caption{Left: absolute value of measurements of the sky-projected spin-orbit angle $\lambda$ for planets we considered in our analysis, differentiated by their status as a compact multi-planet system. Targets from the SLOPE survey are clearly labeled and have larger markers.  Right: distribution of the number of misaligned systems in each population draw from the experiment in Section \ref{sec:statisticalvalidation}.}
    \label{fig:population}
\end{figure*}

We followed a similar methodology to \cite{Radzom2024} and \cite{Rice2022} to perform a statistical test. In 100,000 iterations, we randomly drew 20 planets from the 33 isolated systems without replacement and recorded the number of misaligned planets. This produced a Gaussian distribution representing the expected number of misaligned systems in the compact sample under the null hypothesis that the two populations are drawn from the same distribution. The statistical significance was computed as the offset between this mean and the observed value of one misaligned compact system, normalized by the distribution’s standard deviation. The results are shown in the right panel of Figure~\ref{fig:population}, indicating that isolated systems are more frequently misaligned than compact systems at a significance exceeding 6.1$\,\sigma$. For the purpose of that plot, we use the probabilistic model from \cite{Chen2017} to estimate the mass of TOI-4596~b (we include it only in the Neptune/sub-Neptune class due to its radius). We conclude that the two populations are statistically distinct, with small planets in compact multi-planet systems being preferentially aligned. 

\begin{deluxetable*}{lcccc}
\tablecaption{Summary of the obliquity sample from Section \ref{sec:statisticalvalidation}. \label{tab:statistical_tests}}
\tablehead{
\colhead{Class} &
\colhead{Requirement} &
\colhead{Compact (Aligned)} &
\colhead{Non-Compact (Aligned)} &
\colhead{Significance}
}
\startdata
All Small Planets & $M_p \le 95M_\oplus$ & 20 (19) & 33 (17) & 6.11$\,\sigma$ \\
& & & & \\
Sub-Saturn & $17M_\oplus \le M_p \le 95M_\oplus$ & 8 (8) & 30 (15) & 3.25$\,\sigma$ \\ 
& & & & \\
Neptune/Sub-Neptune &
$5M_\oplus \le M_p \le 50M_\oplus$ & 15 (14)  & 20 (11)  & 5.83$\,\sigma$ \\ 
&\textbf{OR} $2R_\oplus \le R_p \le 6R_\oplus$
& & &  \\
\enddata
\end{deluxetable*}

Repeating the test among sub-populations in the sample confirms that our finding is robust across planetary archetypes. For example, our result is consistent with preliminary findings from \cite{Radzom2024}, although their analysis spanned a narrower range of planet sizes, achieved a lower statistical significance, and counted observations with an arbitrarily high $\lambda$ uncertainty. We repeated our test using the specific mass cuts of those authors, that is, strictly `sub-Saturn' planets with masses $17M_\oplus \le M_p \le 95M_\oplus$. We identified a total of 8 compact systems in which all were aligned, giving a statistical significance of 3.3$\,\sigma$. We repeated the test again for the Neptune/sub-Neptune class\footnote{This class has the largest overlap with the sample from \cite{Polanski2025}, which demonstrated the trend among compact sub-Neptune systems ($R\le4R_\oplus$) with 3.1$\,\sigma$ confidence.} planets, which we defined as either $5M_\oplus \le M_p \le 50M_\oplus$ or $2R_\oplus \le R_p \le 6R_\oplus$, which contained 15 compact systems. For Neptunes and sub-Neptunes, compact multi-planet systems are preferentially aligned compared to other architectures with 5.8$\,\sigma$ confidence. A summary of all three tests is given in Table \ref{tab:statistical_tests}. Although a number of planets smaller than $5M_\oplus$ and $2R_\oplus$ (super-Earth sized) have also had their obliquities measured, almost every one of those systems is compact, preventing us from performing a similar test.

We interpret this test to mean that the processes that act to excite stellar obliquities also disrupt compact systems. If the underlying obliquity excitation mechanisms are predominantly driven by planetary excitation (ZKL oscillations, \citealt{Fabrycky2007,Naoz2011,Naoz2012}; secular chaos, \citealt{Wu2011}; scattering, \citealt{Chatterjee2008}; secular resonance, \citealt{Petrovich2020,Su2025,Handley2026}), the misalignment may be followed by dynamical havoc that separates, engulfs, or ejects planets. If the underlying excitation is driven by primordial misalignment of the stellar spin axis due to a stellar binary or broken disk \citep{Spalding2014,EpsteinMartin2022}, the lack of compact architectures could be simply attributed to differential spin axis precession of individual planets, which disrupts the multi-transiting nature of the system without necessarily decreasing planetary multiplicity. Dedicated RV followup to rule out nearby non-transiting planets might help distinguish these two scenarios.

Our result has important implications for future statistical analyses of this class of planets. That is, samples of spin-orbit angles in isolated and compact architectures should be mixed with caution, as they represent distinct populations. The dynamical origin of major spin-orbit misalignments will be better determined through curated samples of isolated planets.



\subsection{Connection to Disk-Star Misalignments}

\cite{Biddle2025} recently found evidence that as many as one third of planetary systems are born misaligned ($\Delta i\gtrsim10\degree$) from their host star using spatially resolved protoplanetary disks. For a first-order quantitative comparison, we modeled the compact multi-planet systems as a truncated Gaussian distribution:

\begin{equation}
    |\lambda|\sim\mathcal{N}(\mu_\lambda,\sigma_\lambda)|_{[0,\infty)}
\end{equation}
to account for the absolute value of $\lambda$. We used a hierarchical methodology which considered the unique obliquity uncertainty for each system. For this the test, we neglected the HD~3167 system as it is a major outlier that broadens the distribution in a manner inconsistent with the rest of the data. An MCMC exploration of that likelihood indicated that our sample is consistent with a probability density peak at zero degrees ($\mu_\lambda=4.4\degree\pm2.8\degree$) and spread of $\sigma_\lambda=6.3\degree\pm2.7\degree$. Given that we find $\mu_\lambda \sim 0$ (and that all considered systems are consistent with $\lambda\lesssim10\degree$), the sky-projected obliquities of compact multi-planet systems do not show evidence of significant minimum misalignments. Constraints on the true obliquities would be required to confirm if the results are in tension.

We note the additional caveat from \cite{Biddle2025} that disk inclinations are measured at very large scales (10s of au) while these planets orbit at typically $\ll 1$~au. A population of aligned inner disks and misaligned outer disks could also explain why misalignments are not present in our sample.

\subsection{Other Trends}

There is tentative evidence that an orbital dichotomy exists for short period planets with $M_p\sim30 M_\oplus$ (a planet-to-star mass ratio of $\sim10^{-4}$), in which the planets are generally found in an aligned or near-polar configuration \citep{Albrecht2021,Knudstrup2024,EspinozaRetamal2024,Handley2025,Bourrier2025,Rossi2026}. Further, the misalignments appear to be most common among planets occupying the `Neptunian Ridge' coined by \cite{CastroGonzalez2024}, a slight occurrence pileup (compared to nearby dearths of occurrence) for Neptune-like planets orbiting at periods of $\sim$3-6 days. Large-scale surveys which bolster this sample will play an important role in our understanding of both the occurrence and obliquity architectures (our SLOPE survey; the ATREIDES project, \citealt{Bourrier2025}; the POSEIDON survey, \citealt{EpsinozaRetamal2026}). 

Our measurement of TOI-1823~b ($P_\text{orb}=38.8$ days) also ranks among the longest period sub-Saturns for which the obliquity has been measured. Other notable examples include HIP~41378~f (542 days; \citealt{Grouffal2025}), TOI-813~b (84 days; \citealt{Knudstrup2024}), and Kepler-1656~b (32 days; \citealt{Rubenzahl2024a}). The projected obliquities of all these planets are consistent with alignment\footnote{However, we note that the observed stellar rotational periods of HIP~41378 and Kepler-1656 lead the authors to believe the orbits may be misaligned in 3D.}. It may be the case that small planets are preferentially aligned beyond some characteristic distance to the star, similar to the trend observed for warm ($a/R_\star > 11$) Jupiters \citep{Rice2022}. Many more observations of longer-period planets will be necessary to make the distinction.

\section{Conclusion}

We have introduced the new SLOPE survey of stellar obliquity measurements with KPF targeting sub-Saturn and smaller planets with the intention of building a statistically robust sample. To complement our survey, we described a new tool in development for the joint modeling of photometry and RM datasets. With that tool, we analyzed the first four targets from our survey. All four planets---TOI-1386~b, TOI-480~b, TOI-4596~b, and TOI-1823~b---have sky-projected obliquities consistent with alignment. In the process, we confirmed that TOI-4596~b is not a false positive, and encourage future RV followup to determine the planets precise mass. 

We also curated our own sample of small planet obliquities from the literature, and demonstrated that those in compact multi-planet systems are preferentially aligned compared to other architectures with high ($>6\,\sigma$) confidence. Dividing our sample into finer bins of planet size did not reveal any deviations from this trend, although the super-Earth regime is largely unexplored. Experiments that attempt to estimate the relative frequency of spin-orbit excitation processes from the obliquity distribution should be cautious about including compact multi-planet systems. RV followup of misaligned single-transiting systems can help distinguish if they arise from excitation of the planetary orbit or evolution of the stellar spin axis. 

\begin{acknowledgments}

L.B.H. acknowledges support from the National Science Foundation through the Graduate Research Fellowship Program under Grant No. 2139433.  A.W.H. acknowledges funding support from NASA grant No. 80NSSC24K0161. This research was carried out, in part, at the Jet Propulsion Laboratory and the California Institute of Technology under a contract with the National Aeronautics and Space Administration and funded through the President’s and Director’s Research \& Development Fund Program.

The authors acknowledge the use of public TESS data from pipelines at the TESS Science Office and at the TESS SPOC at NASA Ames Research Center. This research has made use of the NASA Exoplanet Archive and Exoplanet Follow-up Observation Program website, which are operated by the California Institute of Technology, under contract with the National Aeronautics and Space Administration under the Exoplanet Exploration Program.

Some of the data presented herein were obtained at the W.\,M.~Keck Observatory, which is operated as a scientific partnership among the California Institute of Technology, the University of California, and the National Aeronautics and Space Administration. We wish to recognize and acknowledge the very significant cultural role and reverence that the summit of Maunakea has always had within the indigenous Hawaiian community. We are most fortunate to have the opportunity to conduct observations from this mountain.

\end{acknowledgments}

\begin{contribution}

\end{contribution}

\facilities{Keck I (KPF), Keck I (HIRES), TESS}

\software{\textsc{astropy} \citep{astropy:2013, astropy:2018, astropy:2022}, 
          \textsc{batman} \citep{batman}, 
          \textsc{emcee} \citep{foremanmackey2013}, 
          \textsc{lightkurve} \citep{lightkurve}, 
          \textsc{matplotlib} \citep{matplotlib}, 
          \textsc{numpy} \citep{numpy}, 
          \textsc{pandas} \citep{pandas}, 
          \textsc{scipy} \citep{scipy}}


\appendix

\section{Transit Timing Variations} \label{sec:ttvs}

\begin{figure}
    \centering
    \includegraphics[width=\linewidth]{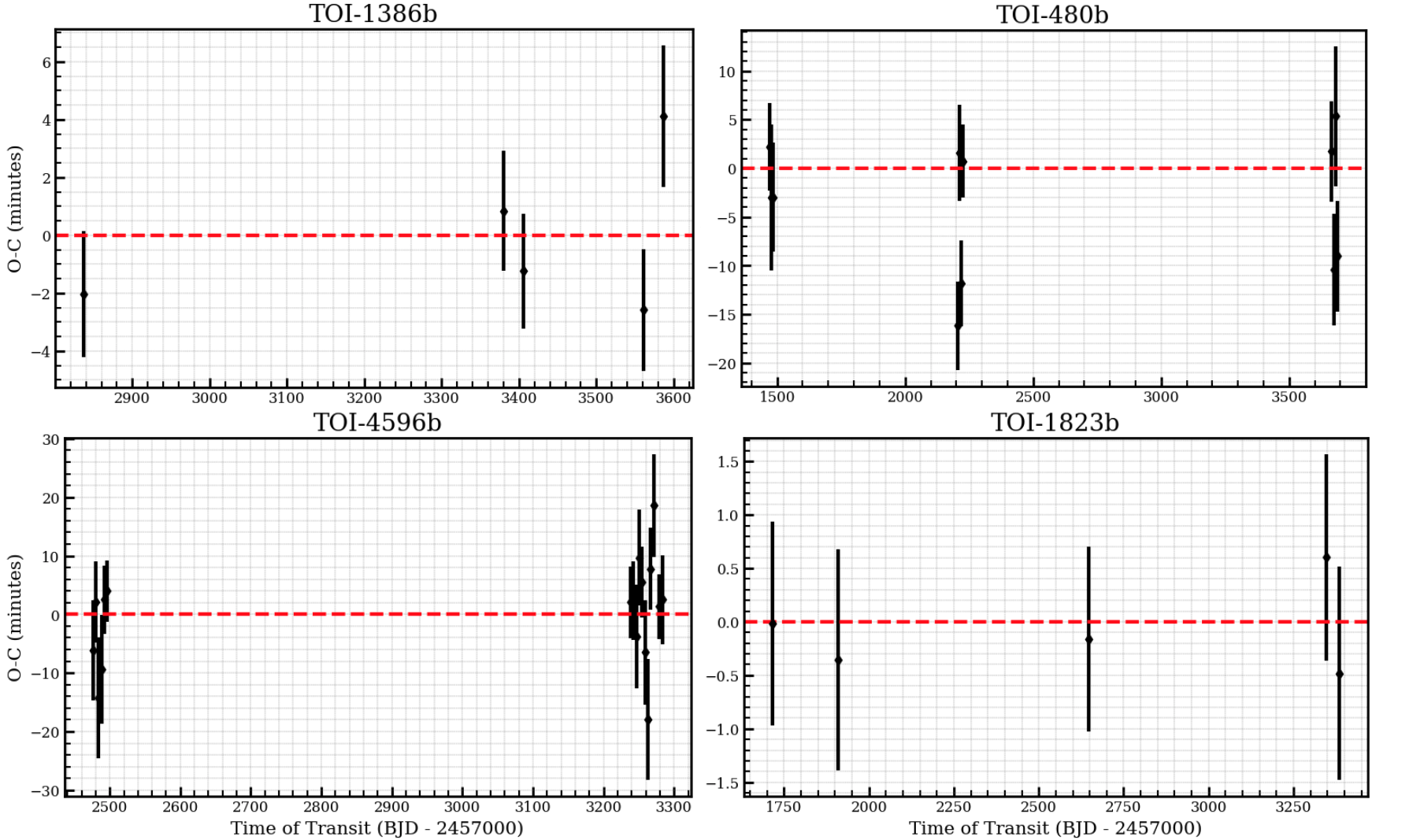}
    \caption{Transit times for all targets after each transit was fit with an independent midpoint (black points). The times are plotted as the best fit observed time minus the computed time (O-C) assuming a linear solution given by the best fit parameters in Section \ref{sec:results}. Uncertainties were estimated using an MCMC approach. None of the targets demonstrate significant deviations from a linear model (red dashed line) which would suggest the presence of additional planets.}
    \label{fig:ttvs}
\end{figure}

To ensure that additional unknown planets did not affect the timing of the transit observed with KPF, we also checked for TTVs. To do this, we fit the TESS photometry (see Section \ref{sec:model}; we used Equation \ref{eq:likelihood} but neglected the KPF observations) globally to get a good inference on the period and calculated the expected transit times in the data. We then fit each individual transit as an independent dataset, where the only free parameter was the timing of the transit. The residuals between the observed and computed transit times from this experiment are shown in Figure \ref{fig:ttvs}; transit times were consistent with no difference from a linear solution for all our targets, indicating TTVs do not influence our dataset. Thus, it can be assured that the transit times are well constrained and ambiguities do not obscure our interpretation of the KPF observations.

\bibliography{main}{}
\bibliographystyle{aasjournalv7}

\end{document}